\documentclass[10pt,journal,compsoc]{IEEEtran}
\ifCLASSOPTIONcompsoc
  \usepackage[nocompress]{cite}
\else
  \usepackage{cite}
\fi

\ifCLASSINFOpdf
\else
\fi
  \usepackage[caption=false,font=footnotesize]{subfig}

\def\MayFloat{true}
\usepackage{pgfplots}
\usepackage{tikz}
\usetikzlibrary{arrows,shadows,backgrounds,calc,decorations.pathreplacing ,decorations.pathmorphing,
positioning
}
\usepackage{float}
\usepackage{colortbl}
\definecolor{ForestGreen}{rgb}{0.0, 0.4, 0.0}
\colorlet{HeadingColor}{ForestGreen}

\colorlet{tableheadcolor}{HeadingColor!25} 
\colorlet{tablerowcolor}{HeadingColor!10} 
\usepackage{url}
\usepackage{xkeyval}	
\usepackage{xcolor}
\definecolor{burntorange}{rgb}{0.8, 0.33, 0.0}
\definecolor{NavyBlue}{rgb}{0.0, 0.0, 0.5}
\definecolor{darkmagenta}{rgb}{0.55, 0.0, 0.55}
\definecolor{ivory}{rgb}{1.0, 1.0, 0.94}
\definecolor{ForestGreen}{rgb}{0.0, 0.4, 0.0}
\colorlet{SeparatorColor}{burntorange}

\usepackage{listingsutf8}
\usepackage{adjustbox}
\usepackage{booktabs}

\colorlet{HeadingColor}{ForestGreen}

	\makeatletter
	\define@key{MEMacros}{color}[green]{\def\ME@color{#1}}
	\define@key{MEMacros}{decorations}{\def\ME@decorations{#1}}
	\define@key{MEMacros}{iconheight}[10pt]{\def\ME@color{#1}}
	\define@key{MEMacros}{language}{\def\ME@language{#1}}
	\define@key{MEMacros}{number}{\def\ME@number{#1}}
	\define@key{MEMacros}{options}{\def\ME@options{#1}}
	\define@key{MEMacros}{plain}[true]{\def\ME@plain{#1}}		
	\define@key{MEMacros}{resize}[true]{\def\ME@resize{#1}}
	\define@key{MEMacros}{shrink}[true]{\def\ME@shrink{#1}}		
	\define@key{MEMacros}{tall}[true]{\def\ME@tall{#1}}
	\define@key{MEMacros}{title}{\def\ME@title{#1}}
	\define@boolkey{MEMacros}{wide}[true]{\def\ME@wide{#1}}
	\define@key{MEMacros}{width}[3cm]{\def\ME@width{#1}}
	\presetkeys{MEMacros}{color=green}{}%
	\presetkeys{MEMacros}{number=1}{}%
	\presetkeys{MEMacros}{plain=false}{}%
	\presetkeys{MEMacros}{resize=false}{}%
	\presetkeys{MEMacros}{shrink=false}{}%
	\presetkeys{MEMacros}{tall=false}{}%
	\presetkeys{MEMacros}{wide=false}{}%
	\presetkeys{MEMacros}{width=3cm}{}%
\makeatother

\ifx\MayFloat\undefined
\else
	\newfloat{program}{thp}{lpp}[section]
	\floatstyle{plaintop}
\fi

\newlength{\mywidth} 

\makeatletter

\newlength{\FigWidth}
\makeatletter
\newcommand\MESourceFile[4][]{
	\setkeys{MEMacros}{wide=false,language={[ANSI]C},options={}, decorations={},#1}%
	\MESetStandardListingFormat
	\ifx\MayFloat\undefined 
		\def\startsource{
			\setlength{\FigWidth}{\textwidth}
		{\color{HeadingColor}
			}
			\par
		}
		\def\stopsource{}
	\else 
		\ifKV@MEMacros@wide 
		   \def\startsource{
			\setlength{\FigWidth}{\textwidth} 
			\begin{program*}[!hbt] 
			}
			\def\stopsource{\end{program*}}			
		\else 
			\if@twocolumn
			  \setlength{\FigWidth}{\columnwidth}
			\else
			  \setlength{\FigWidth}{.7\textwidth}
			 \fi
			\def\startsource{\begin{program}[!hbt]}
			\def\stopsource{\end{program}}
		\fi
	\fi
	\noindent\startsource\hskip-1em
	\vspace{-10pt}
	{
		\mbox{}\phantomsection
		\noindent\begingroup\protected@edef\x{\endgroup\noexpand
			\lstinputlisting[language={\ME@language}, \ME@options, label=#4, name=#4,
			caption ={#3}
			]{#2}}
		\x
		\ME@decorations 
	}					
	\stopsource
}

\makeatletter  
\newcommand\MEtable[6][]{%
	\setkeys{MEMacros}{wide=false,#1}%
	\ifx\MayFloat\undefined
		\def\startsource{
			{\color{HeadingColor}\bfseries\scriptsize #3}\par\vskip-1.8\baselineskip}
		\setlength{\mywidth}{#6\textwidth}
		\def\stopsource{}
	\else
		\ifKV@MEMacros@wide
			\def\startsource{\begin{table*}[!hbtp]	}
			\def\stopsource{\end{table*}}
			\setlength{\mywidth}{#6\textwidth}
		\else
			\setlength{\mywidth}{#6\columnwidth}
			\ifx\WEBBook\undefined\else\setlength{\mywidth}{.5\mywidth}\fi
			\ifx\eBook\undefined\else\setlength{\mywidth}{.7\mywidth}\fi
			\def\startsource{\begin{table}[!hbtp]}
			\def\stopsource{\end{table}}
		\fi
	\fi
	\startsource	  
	\mbox{}\phantomsection 
		\begin{center}
			\vbox{
				\ifx\WEBBook\undefined\else\vskip-1.3\baselineskip\fi
				\begin{minipage}{\mywidth} 			
					\ifx\MayFloat\undefined
					\else{\color{HeadingColor}\caption{#3}\label{#4}}
					\fi 
				\end{minipage}
				\vglue-0.2\baselineskip\par
				\begin{minipage}{\mywidth}
					\ifthenelse{\equal{#5}{}}{}{\vskip-.1\baselineskip{\tiny \copyright #5}	\hfill
						}  
				\end{minipage}
				\par\vskip-.7\baselineskip
				\noindent\makebox[\mywidth]{\color{SeparatorColor}\rule{\mywidth}{1pt}}\par
						\maxsizebox{\mywidth}{.75\textheight}{ 
						#2}
				\par\noindent\makebox[\mywidth]{\color{SeparatorColor}\rule{\mywidth}{1pt}}\par
				\par\vskip-.7\baselineskip
			}
		\end{center}
	\ifx\LecturePrintable\undefined\else\par\vskip-.8\baselineskip\fi
	\stopsource%
}

\makeatother

\ifx\eBook\undefined
	\def\lstsize{\scriptsize}
\else
	\def\lstsize{\tiny}
\fi

\newcommand{\MESetStandardListingFormat}
{
	\lstset{
	    literate=
         {í}{{\'i}}1
         {Í}{{\'I}}1
	    {ő}{{\H{}o}}1
	    {Ő}{{\H{}O}}1
	    {Ű}{{\H{}U}}1
	    {ű}{{\H{}u}}1
	    	}	    
	\lstset{
		numberstyle=\tiny,          
		numbersep=5pt,              
		tabsize=2,                 	 
		inputencoding=utf8/latin2,	
		extendedchars=true,         %
		escapechar=\@,
		breaklines=true,        
		columns=fullflexible,  
		breakatwhitespace=true,    
        keepspaces=true,
        escapeinside={\%*}{*)},          
		frame=tb, 
		framerule=.5pt, 
		rulecolor= \color{SeparatorColor},
		backgroundcolor=\color{ivory},
		basicstyle=\ttfamily\color{black}\normalsize\bfseries, 
		keywordstyle=\bfseries\color{darkmagenta},
		identifierstyle=\bfseries\color{NavyBlue},
		commentstyle=\itshape\bfseries\color{ForestGreen},
		stringstyle=\itshape\bfseries\color{burntorange}, 
		lineskip=0pt,aboveskip=4pt,belowskip=2pt,
		framesep=4pt,rulesep=2pt, 
		showspaces=false,           %
		showtabs=false,             %
		framexleftmargin=0pt,
		framexrightmargin=0pt,
		showstringspaces=false      
	}	
}
\newcommand\MESetListingFormat[2][]	
{
    \MESetStandardListingFormat
    \lstset
    	{
		#1,
		language={#2},		
         numberstyle=\tiny,          
         numbersep=5pt,              
         tabsize=2,                 	 
         inputencoding=utf8/latin2,	
         extendedchars=true,         %
         escapechar=\@,
		breaklines=true,        
		breakatwhitespace=true,    
		escapeinside={\%*}{*)},          
		frame=tb, 
		framerule=.5pt, 
		rulecolor= \color{SeparatorColor},
		 backgroundcolor=\color{ivory},
       basicstyle=\ttfamily\color{black}\lstsize , 
         keywordstyle=\bfseries\color{magenta},
         identifierstyle=\bfseries\color{NavyBlue},
 		commentstyle=\itshape\bfseries\color{ForestGreen},
        stringstyle=\itshape\bfseries\color{burntorange}, 
          lineskip=0pt,aboveskip=4pt,belowskip=2pt,
            framesep=4pt,rulesep=2pt, 
          showspaces=false,           %
         showtabs=false,             %
         xleftmargin=-1pt,
         framexbottommargin=4pt,
         framextopmargin=4pt,
         gobble=5,
		framexleftmargin=0pt,
		framexrightmargin=0pt,
         showstringspaces=false      
	}	
}

\makeatletter
\newcounter{qan}\newcounter{qano}
\newcommand\MEListBalls[3][]{%
	\setkeys{MEMacros}{color=orange,#1}%
	\setkeys{MEMacros}{number=1,#1}%
\setcounter{qan}{0}
\setcounter{qano}{0}
   \begingroup%
    \foreach \x in {#3}
    {  \addtocounter{qan}{1}
    	\addtocounter{qano}{1}
  {\tikz[remember picture,overlay]
    {\expandafter\node[circle, inner sep=2pt, draw,fill=\ME@color,ball color=\ME@color, shading=ball, font=\scriptsize\bfseries, drop shadow]
   at  ([xshift=+10pt,yshift=+2pt]{pic cs:line-#2-\x-end}) {\lstsize\arabic{qan}};\expandafter}}
   }
	\endgroup
}
\makeatother

\lstdefinelanguage
   [x64]{Assembler}     
   [x86masm]{Assembler} 
   {morekeywords={CDQE,CQO,CMPSQ,CMPXCHG16B,JRCXZ,LODSQ,MOVSXD, %
                  POPFQ,PUSHFQ,SCASQ,STOSQ,IRETQ,RDTSCP,SWAPGS, %
                  rax,rdx,rcx,rbx,rsi,rdi,rsp,rbp, %
                  r8,r8d,r8w,r8b,r9,r9d,r9w,r9b}} 

\lstdefinelanguage
   [y86]{Assembler}     
   [x86masm]{Assembler} 
   {morekeywords={ halt, nop, movl, rrmovl, irmovl, rmmovl, mrmovl, rrmovl, jump,%
                  xorl, andl, je, addl, jne, subl, pushl, popl, jXX,%
                  cmovXX, cmovl, cmovle, cmove, cmovne, cmovge, cmovg, return,%
                   rax,rbx,rcx,rdx,xmm0,rbp, load,mulss,%
                   OPl, ZF, SF, OF,%
                   eno, ecc, esv,
                   QCreate, QTCreate, QFCreate, QTerm, QCall, QAlloc,
                   QWait, QPWait, QIWait, QInt
                  },
      morecomment=[l]{\#},
      sensitive=false,
     } 

		\usepackage[toc,acronym]{glossaries}

		\makeglossaries

  \newacronym{ALU}{ALU}{Arithmetic and Logic Unit}
  \newacronym{CMP}{CMP}{Chip Multi Processor}
  \newacronym{CPU}{CPU}{Central Processing Unit}
  \newacronym{CSTB}{CSTB}{Computer Science and Telecommunications Board}
  \newacronym{EDVAC}{EDVAC}{Electronic Discrete Variable Computer}
  \newacronym{ENIAC}{ENIAC}{Electronic Numerical Integrator and Computer}
  \newacronym{EMPA}{EMPA}{Explicitly Many-Processor Approach}
  \newacronym{FPGA}{FPGA}{Field Programmable Gate Array}
  
  \newacronym{GPGPU}{GPGPU}{General-Purpose Graphics Processing Unit}
  \newacronym{HW}{HW}{hardware   }
  \newacronym[description={Instruction Level Parallelism}]{ILP}{ILP}{Instruction Level Parallelism}
  \newacronym{I/O}{I/O}{input/output}
  \newacronym[description={Instruction Set Architecture}]{ISA}{ISA}{Instruction Set Architecture}
  \newacronym{LU}{LU}{logic unit}
  
\newacronym[description={Million Instructions Per Second}]
{MIPS}{MIPS}{Million Instructions Per Second}
 \newacronym[description={Object Oriented Programming}]{OOP}{OOP}
 {Object Oriented Programming}

  \newacronym[description={Operating System}]{OS}{OS}{Operating System}
  \newacronym[description={Program Counter}]{PC}{PC}{Program Counter}
  \newacronym{PRAM}{PRAM}{Parallel Random Access Model \protect{\cite{Vishkin:FineGrainedProgramming} }
  }
  \newacronym{PU}{PU}{Processing Unit}
  \newacronym{QT}{QT}{Quasi-Thread}
  \newacronym{RC}{RC}{reconfigurable}
  \newacronym{RT}{RT}{real time}
  \newacronym{SIMD}{SIMD}{Single Instruction Multiple Data}
  \newacronym{SPA}{SPA}{Single Processor Approach}
  \newacronym{SV}{SV}{supervisor}
  \newacronym{SW}{SW}{software}
  \newacronym[description={Thread Level Parallelism}]{TLP}{TLP}{Thread Level Parallelism}

\newacronym{XMT}{XMT}{eXplicit Multi-Threading}

\usepackage{listings}

\makeatletter
\newcommand{\highlighted}[2][]{%
	\setkeys{MEMacros}{color=darkmagenta!40!black!80,#1}%
	{\textit{\textbf
			{\textcolor{\expandafter\ME@color\expandafter}{#2}}}}%
}

\newcommand\MEfigure[6][]{
	\setkeys{MEMacros}{wide=false,#1}%
	\ifx\MayFloat\undefined
		\def\startsource{ 
		}
		\setlength{\mywidth}{#6\textwidth}
		\def\stopsource{}
	\else
		\ifKV@MEMacros@wide
			\def\startsource{\par\begin{figure*}[!hbt]	
					}
			\def\stopsource{\end{figure*}}
			\setlength{\mywidth}{#6\textwidth}
		\else
			\setlength{\mywidth}{#6\columnwidth}
			\ifx\WEBBook\undefined\else\setlength{\mywidth}{.55\mywidth}\fi
			\ifx\eBook\undefined\else\setlength{\mywidth}{.7\mywidth}\fi
			\def\startsource{\par\begin{figure}[!hbtp]}
			\def\stopsource{\end{figure}}
		\fi
	\fi
	
	\par\startsource	  
	\mbox{}\phantomsection 
	{{\centering
	
			\vbox{
				\ifx\WEBBook\undefined\else\vskip-.3\baselineskip\fi
				\begin{minipage}{\mywidth}
					\ifthenelse{\equal{#5}{}}{}{\vskip-.1\baselineskip{\tiny \copyright #5}	\hfill
						}  
				\end{minipage}
				\par\vskip-.9\baselineskip
				\noindent\makebox[\mywidth]{\color{SeparatorColor}\rule{\mywidth}{1pt}}\par\vskip.2\baselineskip
						\maxsizebox{\mywidth}{.75\textheight}
						{ 

						\includegraphics[width=\mywidth,keepaspectratio]{#2}
						}
					\par\vskip-.8\baselineskip\par
				\noindent\makebox[\mywidth]{\color{SeparatorColor}\rule{\mywidth}{1pt}}\par
			
				\begin{minipage}{\mywidth} 			
					\ifx\MayFloat\undefined
						{\scriptsize {\color{HeadingColor}#3}}
					\else{\color{HeadingColor}\caption{{#3}}\label{#4}}
					\fi 
				\end{minipage}
			}
		
	\ifx\LecturePrintable\undefined\else\par\vspace{-15pt}\fi				
}}
	\stopsource%
}

\newcommand\MEtikzfigure[6][]{
	\setkeys{MEMacros}{wide=false,#1}%
	\ifx\MayFloat\undefined
		\def\startsource{ 
		}
		\setlength{\mywidth}{#6\textwidth}
		\def\stopsource{}
	\else
		\ifKV@MEMacros@wide
			\def\startsource{\par\begin{figure*}[!hbt]	
					}
			\def\stopsource{\end{figure*}}
			\setlength{\mywidth}{#6\textwidth}
		\else
			\setlength{\mywidth}{#6\columnwidth}
			\ifx\WEBBook\undefined\else\setlength{\mywidth}{.55\mywidth}\fi
			\ifx\eBook\undefined\else\setlength{\mywidth}{.7\mywidth}\fi
			\def\startsource{\par\begin{figure}[!hbtp]}
			\def\stopsource{\end{figure}}
		\fi
	\fi
	
	\par\startsource	  
	\mbox{}\phantomsection 
	{{\centering
	
			\vbox{
				\ifx\WEBBook\undefined\else\vskip-.3\baselineskip\fi
				\begin{minipage}{\mywidth}
					\ifthenelse{\equal{#5}{}}{}{\vskip-.1\baselineskip{\tiny \copyright #5}	\hfill
						}  
				\end{minipage}
				\par\vskip-.9\baselineskip
				\noindent\makebox[\mywidth]{\color{SeparatorColor}\rule{\mywidth}{1pt}}\par\vskip.2\baselineskip
						\maxsizebox{\mywidth}{.75\textheight}
						{ 
						#2
						}
				\noindent\makebox[\mywidth]{\color{SeparatorColor}\rule{\mywidth}{1pt}}\par
			
				\begin{minipage}{\mywidth} 			
					\ifx\MayFloat\undefined
						{\scriptsize {\color{HeadingColor}#3}}
					\else{\color{HeadingColor}\caption{{#3}}\label{#4}}
					\fi 
				\end{minipage}
			}
		
	\ifx\LecturePrintable\undefined\else\par\vspace{-15pt}\fi				
}}
	\stopsource%
}

\makeatother 

\begin{document}
\title{A configurable accelerator for manycores:\\the Explicitly Many-Processor Approach}
%
%
%
\author{J\'anos V\'egh
\IEEEcompsocitemizethanks{\IEEEcompsocthanksitem J. V\'egh is with Faculty of Mechanical Engineering and Informatics, University of Miskolc, Hungary\protect\\
E-mail: J.Vegh@uni-miskolc.hu
}
\thanks{Manuscript received July 10, 2016; revised August 26, 2016.}}

\markboth{The Explicitly Many-Processor Approach,~Vol.~14, No.~8, August~2016}%
{V~\egh: the Explicitly Many-Processor Approach}
%



\IEEEtitleabstractindextext{%
\begin{abstract}
A new approach to designing processor accelerators is presented.
A new computing model and a special kind of accelerator with dynamic 
(end-user programmable) architecture 
is suggested. The new model considers a processor, in which a newly introduced
supervisor layer coordinates the job of the cores.
The cores have the ability
(based on the parallelization information provided by the compiler, and
using the help of the supervisor) to outsource part of the job they received
to some neighbouring core. The introduced changes essentially and advantageously
modify the architecture and operation of the computing systems.
 The computing throughput drastically increases,
 the efficiency of the technological implementation (computing performance per logic gates) increases, the non-payload activity for using operating system services decreases,
the real-time behavior changes advantageously, and connecting accelerators
to the processor greatly simplifies.
Here only some details of the architecture and operation
of the processor are discussed, the rest is described elsewhere.

\end{abstract}

\begin{IEEEkeywords}
computer architecture, processor accelerator, manycore processor, many-processor approach
\end{IEEEkeywords}}

\maketitle
\lstset{language={[x86masm]Assembler}, basicstyle=\ttfamily\color{black}\normalsize}

\IEEEdisplaynontitleabstractindextext

\IEEEpeerreviewmaketitle

\IEEEraisesectionheading{\section{Introduction}\label{sec:introduction}}

%
\IEEEPARstart{A}{bout} a decade ago, "\emph{growth
in single-processor performance has stalled -- or at best
is being increased only marginally over time}"~\cite{ComputingPerformance:2011}.
Since the computing industry heavily influences
virtually all industrial segments, the stalling crashed the long-term forecasts in the industry, economy, etc.
The need for growing performance triggered research in many directions, 
from "rebooting computing"~\cite{RebootingComputing2013} through 
using a "cross-layer approach"~\cite{HSWscalable2012} to using accelerators
combining different technologies, like reconfigurable \gls{GPGPU}~\cite{Braak:2016:RRG:2899032.2890506}.
Especially the popular and powerful many-core processors provide huge computing capacity
and sever problems, like
"\textit{multicore and manycore 
vendors and runtime systems cannot possibly support
the virtually unlimited number of processor configurations"}~\cite{SHIMM13:Multicore}.
It is not yet checked, however, what hidden reserves for acceleration can be disclosed 
in the operation of the conventional processors itself.

In addition to the \gls{HW} issues, also the \gls{SW} makes its contribution:
"\textit{parallel programs ... are notoriously difficult to write, 
test, analyze, debug, and verify, much more so than 
the sequential versions}"~\cite{ReliableParallel2014},
 and doubts like
"\textit{Chip multi-processors have emerged as one of the most effective uses of the huge
number of transistors available today and in the future,
but questions remain as to the best way to leverage \gls{CMP}s
to accelerate single threaded applications}"~\cite{HWcontrolledthreadsMahesri:2007}.
This is why "\textit{cross-layer approach spanning from hardware to user-
facing software is necessary to successfully address this
problem}"~\cite{Adaptiveresourcecontrolinmulticoresystems2013}.

In this paper a "better way" of accelerating  single threaded applications is searched,
first scrutinizing the hidden reserves  in the operation of the stored program processor based computing systems.
During this, it is shown that the final reason of the present stalling
(in addition to the already known reason: the finite speed of the light)
is the 70-years old single-processor approach, which dominates both computer
construction and programming. The introduced \gls{EMPA} viewpoint solves
some old problems of computing. It increases the computing troughput, depending on 
the context up to several dozens times higher, allows to build more deterministic real-time systems, etc. At the same time, it allows to simplify the internal architecture, 
to use less transistors for the chips.

\section{The single-processor approach}\label{sec:singleprooc}

According to the state of the art of his time, Neumann considered only one processor
and formulated the principles of operating a processor considering only the execution
of a single machine instruction. Considering his paradigms carefully,
neither of them contradicts to the present many-processor approach
(see section \ref{sec:manyproc}).

\subsection{Lack of time dependence}

The \emph{roots of the event-oriented processing}
were implicitly present in Neumann's approach: the next instruction can only be executed when the
processor is not any more busy with executing the current instruction.
The time, however, in the paradigms is considered as an implementation detail.
Although later the lack of considering the time explicitly
was identified as one of the fundamental issues in computing 
\cite{IannucciIssues:1988}, handling time (as well as synchronizing)
became the task of operating systems, which they can solve in
a quite resource-wasting way \cite{SynchronizationEverything2013}.

The availability of the processing unit,
the instruction and the data are critical factors, and all they have
a timely behavior. The most successful
approaches to improving performance of processors modified the conditions of processor availability (methods for  reducing the 
instruction and data memory access time are not touched here).
The pipelining separated the signal into "ready to accept data" and
"result is available" signals, the hyper-threading connected the ready-to-run thread to the processing unit.
Critics like "\emph{HT generally improves processor resource utilization efficiency,
but does not necessarily translate into overall application performance gain}" \cite{NASAhyper2011}, call  \textit{to scrutinize 
the processor availability condition and cross-layer functionality}, if one wants to improve single-processor performance
using several processing units.

\subsection{Atomic processing unit}

Hyper-threading separates 
the hardware accelerators (like registers and core-level cache)
from the raw processing power, which is a clear recognition of the 
fact that from programming point of view, the single machine instruction
(a \gls{HW} unit) is too small (cannot make reasonable functionality),
while a complete process (a \gls{SW} unit)
is too big (wastes time with waiting),
so some intermediate unit: \gls{HW}-supported \gls{QT}
should be introduced. This suggests to check, \textit{what the optimum
size of the right unit is, and how it should be supported from \gls{HW} and \gls{SW}}. 
Also it shows that \textit{the effective problem solution can only be
reached through \gls{HW}/\gls{SW} codesign, i.e through crossing \gls{HW}/\gls{SW}
layers~\cite{Adaptiveresourcecontrolinmulticoresystems2013}}.

\subsection{Multiple processing units}

From again another side, from the simple out-of-order processing through
multiple arithmetic units to speculative evaluation,
several kinds of \emph{hidden} processing units have been introduced 
and they successfully accelerated the operation of the processor.
There is no question, that more processing units are needed
to make parallel operations (and in this way  apparently the processing) quicker.
However, those solutions forget about one of the most important principles
of Neumann: \emph{computer should be simple}.

In order to achive higher performance, "\emph{computers have thus far achieved this goal
at the expense of tremendous hardware complexity --
a complexity that has grown so large as to challenge
the industry's ability to deliver ever-higher performance}" \dots and
\dots "\emph{the ever- increasing complexity of superscalar processors would have a
negative impact upon their clock rate, eventually leading to a leveling off of the rate
of increase in microprocessor performance}" \cite{EPIC:2000}.
This suggests to check \textit{what the optimum way of using
multiple processing units is}, without wasting computing and electric power?

\subsection{Multitasking issues}

Different reasons directed the designers to share the computing
capacity and other resources between different tasks, starting with the 
age of single-processor systems. The methods, however, have 
been prolonged to  the era, when several processing units could be 
used. Even today, the external peripherals \textit{interrupt} the control flow
(although some other core could do interrupt servicing), in
the many-processor systems all \gls{PU}s are \textit{central}, the
operating systems providing services and scheduling the operation
of the running tasks take the processor time from the payload jobs, 
changing context between user and kernel modes causes a considerable
non-payload activity, the hardware scheduling makes the software
operation non predictable, etc.

A related special issue for accelerators is that an accelerator
is always outside the processor, and it is efficient because
it works differently from the conventional, programmable processors. 
Several problems must be solved in order to connect 
the stored program (von Neumann) processors with the rest of the world.
The processor only offers \gls{I/O} bus for connecting an accelerator
to the processor. However, in the today's environment an \gls{OS} must
provide protection (virtualization) for the \gls{I/O} operations.
It is only possible in protected mode, and the context change from and back to user mode is 
extremely expensive: it is in the range of dozens of thousands clock periods
for the modern \gls{HW} architectures and \gls{OS}s~\cite{hallaron}.
This fact \textit{restricts the utilization of general-purpose accelerators 
to accelerate only activities long enough to be not disproportional with that 
offset time}.

\section{The many-processor approach}\label{sec:manyproc}
The principle of operation of stored program processors
(see Fig~\ref{fig:CompareComputerPrinciples}(a)) and its engineering implementation
(see Fig~\ref{fig:CompareComputerPrinciples}(b)) became
quite different during the time, due to the efforts to enhance
processing speed of the computer. Some of the enhancements are none-essential:
although without cache memories the operation of the processor
would be painfully slow, the processor would work. 
Also, the highly successful accelerators, the internal registers,
are not strictly needed for the operation. 
The processors could work with a strongly limited number of registers \cite{Mahlke:1992:LimitedRegisters}, or as the example of the very first
EDVACs prove \cite{FateofEDVAC1993}, even without registers. Also recall, that it is advantageous to separate registers (as "glue" material, together with internal state and cache) from the processing unit, see hyper-threading or shadow register set at some interrupt servicing.

\MEtikzfigure[resize]{
	\centering
	\subfloat[Principle of the computation in stored program computers \label{fig:ComputerPrinciple}]{
		\begin{tikzpicture}[scale=1.2]
		{
			\shade[ball color=magenta,shading=ball, font=\scriptsize\bfseries
			]
			(4,1.5) ellipse (1.55 and .47);  
			\node[color = white] at (4.1,1.5) {\Large \bf Processing};
		}
		{
			\shade[rotate=90,right color=green!0,left color=green!70](2.95,-2.9) ellipse (1.1 and .3);
			\node[rotate=90] at (2.9,2.6) {Data};
		}
		{
			\shade[rotate=90,right color=green!70,left color=green!0]	(2.95,-5.1) ellipse (1.1 and .3);
			\node[rotate=90] at (5.1,2.8) {Instruction};
		}
		{
			\shade[bottom color=blue!00,top color=blue!70] (4,3.83) ellipse (1.3 and .4);
			\node at (4.0,3.83) {Memory};
		}
		
		{
			\shade[bottom color=blue!00,top color=blue!70] 
			(4,.6) ellipse (1.2 and .5);  
			\node at (4.0,0.6) {Processing unit};
		}
		\phantom{
			\shade[rotate=90,bottom color=yellow!0,top color=yellow!70]
			(-0.5,-3.3) ellipse (0.7 and .3);  
			\node[rotate=90] at (3.3,-0.3) {Registers};
		}
		\end{tikzpicture}
	}
	\hskip 2mm
	\subfloat[Engineering implementation of stored program computers \label{fig:ComputerEngineering}]{
		
		\begin{tikzpicture}[scale=1.2]
		{
			\shade[bottom color=yellow!70,top color=yellow!0]
			(4,2.1) ellipse (1.1 and .4);  
			\node[font=\bfseries] at (4.0,2.1) {Cache};
		}
		{
			\shade[ball color=magenta,shading=ball, font=\scriptsize\bfseries
			]
			(4,1.5) ellipse (1.55 and .47);  
			\node[color = white] at (4.1,1.5) {\Large \bf Processing};
		}
		{
			\shade[rotate=90,right color=green!0,left color=green!70](2.95,-2.9) ellipse (1.1 and .3);
			\node[rotate=90] at (2.9,2.6) {Data};
		}
		{
			\shade[rotate=90,right color=green!70,left color=green!0]	(2.95,-5.1) ellipse (1.1 and .3);
			\node[rotate=90] at (5.1,2.8) {Instruction};
		}
		{
			\shade[bottom color=blue!00,top color=blue!70] (4,3.83) ellipse (1.3 and .4);
			\node at (4.0,3.83) {Memory};
		}
		
		{
			\shade[bottom color=blue!00,top color=blue!70] 
			(4,.6) ellipse (1.2 and .5);  
			\node at (4.0,0.6) {Processing unit};
		}
		{
			\shade[rotate=90,bottom color=yellow!0,top color=yellow!70]
			(-0.2,-3.3) ellipse (0.7 and .3);  
			\node[rotate=90] at (3.3,-0.2) {Registers1};
		}
		{
			\shade[rotate=90,bottom color=yellow!0,top color=yellow!70]
			(-0.3,-3.9) ellipse (0.7 and .3);  
			\node[rotate=90] at (3.9,-0.0) {Cache};
		}
		{
			\shade[rotate=90,bottom color=yellow!0,top color=yellow!70]
			(-0.2,-4.5) ellipse (0.7 and .3);  
			\node[rotate=90] at (4.5,-0.2) {Registers2};
		}
		
		{
			\draw[rotate=0,ball color=red,shading=ball, font=\scriptsize\bfseries] (2.4,1.8) ellipse (1.0 and .3);  
			\node[color=white]  at (2.45,1.8) {\bf Data avail};
		}
		
		{
			\draw[rotate=0,ball color=red,shading=ball, font=\scriptsize\bfseries] (5.6,1.8) ellipse (1.0 and .3);  
			\node[color=white] at (5.6,1.8) {\bf Instr avail};
		}
		{
			\draw[rotate=0,ball color=red,shading=ball, font=\scriptsize\bfseries] (4,1.05) ellipse (1.1 and .3);  
			\node[color=white] at (4.0,1.05) {\bf ALU available};
		}
		\end{tikzpicture}		
	}
}	
{Comparing the theoretical and engineering operation of computers}
	{fig:CompareComputerPrinciples}{}{}
The many-core processors provide an ideal field to implement 
an \gls{EMPA} environment.
The idea here is \textit{to provide the cores with the ability to outsource
part of the job they received}. I.e. they remain responsible for doing the job,
but not necessarily "personally". 
The approach mathematically covered by the theory of "communicating serial processes"~\cite{CommunicatingSerialProcesses:2015}
and some additional aspects are discussed in~\cite{ManyCoreAwareVegh2014}.

It requires of course a lot of changes:
some external logic must concert the work of cores (data transfer, dependency 
and synchronization), the information about the possible outsourcing must be
prepared at compile time rather than at runtime, the code must be cut to optimally 
sized, partly independent \gls{QT}s, the processor must be notified about
the pre-calculated parallelization possibilities, etc.
The \gls{QT}s can be embedded into each other, so the core receiving 
some outsourced \gls{QT} can also attempt to find helper cores\footnote{Since in any given moment there is a one-to-one correspondance 
between an allocated core and the \gls{QT} running on it,
these two terms are used interchangably: they emphasize the \gls{HW} and the 
\gls{SW} side of the same thing, similarly as the processor and the 
process do in single-processor approach.} to make further outsourcing.

\subsection{Availability of processing unit}
In the Neumann paradigms,
the task of the control unit is only to provide a "proper sequencing".
In the single-processor approach, the control unit allows sequential stepping,
jumping, calling and returning, and even interrupt servicing. 

In many-processor approach, it is also possible to report the
availability of several processing units (this time explicitly,
unlike hidden processing units in the single-processor approach).
For compatibility,
a new control layer is introduced. The \gls{SV} is a second
control layer in the many-core processor above the cores, which (among other functionalities)
partly takes care of providing this availability signal. Since it knows 
about all cores in the processor, it can provide 'ready' signal
as long as at least one core is ready to process instructions.
In this way the processor is able to receive new instructions as long as
at least one of the cores is ready to work.  

\subsection{Atomic units}
Unfortunately, only absolutely independent machine instructions 
can be distributed between \gls{PU}s in this way: the "glue" carries valuable state 
information between individual machine instructions.
For allowing  this internal state transfer, the code must be cut into \gls{QT}s
in a reasonable way.
However, to select the proper \textit{granularity} is hard.

As mentioned, the machine instructions are (in general) too small to form a \gls{QT}.
The ideal way would be to use the \gls{PU}s as stateless automatons, which would result in too coarse granularity,
like threads. To allow for a more effective \gls{QT} size, a \textit{quasi-stateless
automaton model} is used. Upon beginning to execute a \gls{QT},
a full-fledged "glue" set, comprising register file, flags (clone) and cache (shared),
is provided for the newly hired \gls{PU},
and a limited amount of "glue" can be returned
in a (by the supervisor) synchronized way when a \gls{QT} is finished.
Since \gls{QT}s are running on the cores in a separated way,
for providing atomic units, 
using QTs is ideal: when the core returns 'ready' signal, the action 
protected by the \gls{QT} is finished: both 'owner' and 'others' must wait it.

\subsection{Mapping QTs to cores}
As the \gls{QT}s can create other (child) \gls{QT}s, they form a kind of processing \textit{graph}.
A virtually infinite number of graph nodes must be mapped to a finite
number of physical cores, so sometimes the new \gls{QT}s must wait for computing resource.
The supervisor's core allocation algorithm prefers reaching
the leaves of the graph rather than opening new forks, and as 
an emergency mechanism for the case when \gls{SV} runs out of available cores, the cores can suspend processing 
their own \gls{QT}s, borrowing their own resources to their
child-\gls{QT}s while they are executed.

Another crucial question is how the at compile time labelled \gls{QT}s
can refer to the \gls{QT}s distributed at run time to cores, having "random" availability.
The \gls{QT}s wanting to communicate with other \gls{QT}s refer to the compile-time
address of the \gls{QT} and the \gls{SV} translates this address to a runtime physical 
core number. Because of this, \gls{QT}s have data structures similar to the processes
running under \gls{OS}s.

\subsection{Synchronizing cores}
The supervisor maintains configurable\footnote{
This feature is between programmable, since it takes the configuration information from the object file and reconfigurable, because it changes
the type of connection between the components inside the processor}
 parent-child relationships between the cores,
so the cores have information whether a \gls{QT} is running, either in their
own scope, or in scope of their parent. Without explicit syncronization, the cores work in parallel on different \gls{QT}s
of the originally single-threaded code. This kind of synchronization
is done at processor level, in one clock cycle; rather then the
conventional methods, which need awfully long times \cite{YavitsMulticoreAmdahl2014}.
Even, since waiting is handled by the \gls{SV} based on signals handled by the \gls{SV},
no time is used when there is no need to wait.

\subsection{Linking cores}
As contented by Amdahl \cite{AmdahlSingleProcessor67}, to increase further
the performance, some cooperation between cores is necessary. In \gls{EMPA} it is
implemented in such a way that the cores pass the control signals and data to the \gls{SV},
which -- like a telephone switching center the calls -- forwards them
to the right core.
The \gls{SV} plays a key role here: it handles all resources
of the processor. It connects and synchronizes
the cores  and delivers the limited amount of information,
and even, it provides partners when one of the cores signals that
it could advantageously use more processing capacity.
Also, it concerts collective processing and takes over some functionalities from the \gls{OS},
which functionalities are traditionally missing from processors. 

When creating a new child \gls{QT}, the new core
must be able to continue the processing from the point where its parent is,
i.e. the "glue" of the parent must be \textit{cloned} (using dedicated wiring) to the child.
 Also, upon termination,
the child might send back the content of the link register (clone back, using similar method).
The cores are working in parallel, so a special syncronization mechanism
must be used. The data intended to be sent to the other party, are latched by the sender (see also Figs.~\ref{fig:EMPAparentchild} and~\ref{fig:TwoLevelDiagram}).
Later some triggering signal transfers the data to its destination,
where they are latched again, and the cores need explicitly use the latch register
to access its content. In this way all partners send data to and receive data from their partner
when they need it, providing a proper syncronization.
The cores see those transfers as using pseudo-registers~(see section~\ref{sec:pseudoregisters}).

\subsection{Subroutines, interrupts, traps}
A main task related to subroutines is to remember the return address.
In the \gls{EMPA} way, starting a \gls{QT} means providing also a new core for the
\gls{QT}, and the processing in the parent code continues immediately,
at the address next to the \gls{QT}, i.e. at the return address.
That means, that the "return address" 
shall only be stored for the time of creating a QT (one clock period)
and it is automatically remembered by the \gls{SV}.
So, a \gls{QT} behaves very similarly to a subroutine call,
but the \gls{QT} itself is embedded in the "calling" code flow.
Implementing a special metainstruction for subroutine call just allows
to place the body of the subroutine outside the main code flow.

In the single-processor approach, when another process is to be run,
the processor must be stolen
from the running process, which needs a lot of saving and restoring,
as well as context changing in advanced systems; i.e. rather much
non-payload activity. In \gls{EMPA} approach, a core can be reserved
for interrupt servicing. It can be prepared (even in kernel mode)
and waiting for the interrupt. When the interrupt arrives 
to the core waiting in power economy mode, it immediately starts its servicing,
without any duty to save and restore, saving processing time and memory cycles.
Note that \textit{in \gls{EMPA} approach no context change is needed,
resulting in several hundreds of performance gain} relative to the conventional 
handling.

Similarly, cores can be prepared to provide kernel services.
Some system services, for example semaphore handling, do not really need
all the facilities of the \gls{OS}, they can be implemented in some alternative way.
As our former measurements on soft system~\cite{Vegh:2014:ICSOFTsemaphore} proved,
such alternative implementation resulted in performance gain about 30,
although in that case no context changing was needed.
Similar gain can be expected when implementing \gls{OS} services with \gls{EMPA}.
The gain factor will surely be increased because of the eliminated 
context change, but the concrete gain will depend on the functionality of the service.
In addition to making context changing
(in both directions) obsolete, the kernel and user codes can run even partly parallel.

\subsection{Mass processing}

The physical vicinity also allows to implement certain kinds of
cooperation between the cores. 
Here the source of performance gain can be to eliminate
control-type instructions (like loop counter advancing, checking, jumping)
as well as to eliminate unnecessary stages of instruction execution,
like 
read and writeback when only the final result is interesting.

A typical example is summing up elements of a vector (see Listing~\ref{lst:Qasum04}): the read and write back of the partial sum is needed only when considering the machine instructions 
as atomic unit. In mass operating mode, 
the parent can sum up summands provided by its children, 
in frame of a machine instruction.

\subsection{Linking special accelerator}

As it follows from the description above, for the \gls{SV} a core is represented
as a source and destination of signals and data.
i.e. an extremely simple interface is provided for linking \gls{QT}s.
Since the \gls{QT} can be as large as a \gls{SW} thread, 
and \gls{SV} only "sees" the signals and data, but no \gls{HW} at all,
\gls{EMPA} provides an 
extremely simple interface for linking any kind of external accelerator
(without any non-payload activity) or even \gls{HW}-implemented \gls{SW} processes
like~\cite{borph2000}.

\begin{figure*}
\includegraphics[width=\textwidth, keepaspectratio]{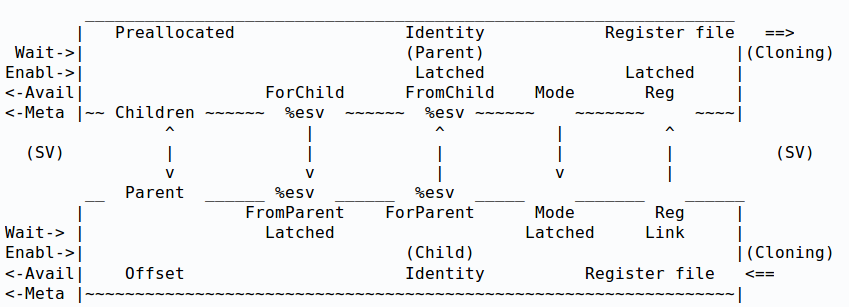}
\caption{The communication signals and data between parent and child cores in EMPA}
\label{fig:EMPAparentchild}
\end{figure*}

\section{Architecture of an EMPA processor} \label{sec:empaarchitecture}

Many of the experts of the field expect some kind of solution
from the \gls{RC} technology (see for example~\cite{Hartenstein07} and its cited references).
However, since the computing density~\cite{ComputingDensity2008} cannot be increased using this 
technology, the \gls{RC} elements are mostly used to control the 
conventional components of the processor (for example \cite{Epiphany2014, Braak:2016:RRG:2899032.2890506})
or to accelerate some \gls{OS} services~\cite{Akesson01}, or one can find even complete
\gls{RT} \gls{OS}s (for example~\cite{Ferreira09}) implemented in \gls{HW}.
At the other 
end of the scale one finds also operating systems
with \gls{SW} threads implemented in \gls{HW}~\cite{borph2000}, too.
When thinking about the \gls{EMPA} architecture, one should have in mind also
\gls{RC} technology combined with the conventional processor technology.
An \gls{EMPA} processor is akin an end-user 
configured dynamic (processor) architecture using dedicated wiring and 
special \gls{LU}s (cores). 

\subsection{Using multiple processing units, explicitly}

Of course, the introduced new operating principle requires modifications also
to the internals of the processor. Since our goal (in addition to
boost single-thread performance) is to preserve as much compatibility
as possible, with the conventional computing,
some components are greatly similar to the conventional ones,
and also the newly introduced components have functionalities,
which can be formulated using terms similar to the conventional computing.

\subsubsection{The processor}
For the outside world, the processor is nearly unchanged.
It receives a stream of instruction codes, which is enriched
with metainstructions, describing the suggestions of the compiler
about increasing performance. Internally, however, it works
differently.

\subsubsection{The cores}
The cores in an EMPA processor are mostly similar to the present single-core
processor, with some extra functionality. 
The cores provide and receive extra signals (see also see also Fig.~\ref{fig:EMPAparentchild}) for/from the \gls{SV}

\begin{itemize}
\item \highlighted{Availability} a core is available when it is not executing a code chunk, not preallocated for a future task,
\index{preallocation}
and not disabled for some reason (like overheating)
\item \highlighted{Enabling} when a core is rented, it gets enabled and disabled again
when the code chunk terminates
\item \highlighted{Waiting} a core can be blocked when waiting for 
the termination of some \gls{QT} running on another core
\item \highlighted{Metainstruction} signals that during pre-fetch a metainstruction found
\end{itemize}

Also store the cores extra information (see also Fig.~\ref{fig:EMPAparentchild} and section~\ref{sec:parentchild}):
\begin{itemize}
\item \highlighted{Identity} The cores are identified by a (hard) "one-hot" bitmask
\item \highlighted{Parent} The (configurable) identifying bit mask of the parent of the core
\item \highlighted{Children} The (configurable) ORed value of the bitmasks of cores with \gls{QT} created by the \gls{QT} running on this core
\item\highlighted{Preallocated} The (configurable) ORed value of the bitmasks  of cores preallocated for this core
\item\highlighted{Offset} The (configurable) memory address of the 
\gls{QT} the core runs
\end{itemize} 

\subsubsection{The supervisor}
This new unit is a second, end-user configurable control layer,
 which, among others, handles the computing resources
 and reports "ALU avail" signal.
It can be considered as a kind of second-level control unit, 
implemented as a control layer above and between the \gls{PU}s (cores). 
It transfers signals to and from cores; as such, all data and control transfers can be implemented between \gls{SV} and \gls{PU}, so
a 'star' topology shall be implemented, and there is no need 
to wire the cores with all other cores.

The  \gls{SV} is responsible for providing a new processing unit
(to rent a core) when needed, to  handle properly the termination signals from the 
rented cores (including handling the mentioned bitmasks), to provide synchronization signals for a core running
a \gls{QT}, provide (limited) synchronized data connection between the cores, etc.
 It  comprises several simple,
easy to implement and quick to execute  functionalities.

The \gls{SV} is responsible for all resources,
so  it can only be used in a sequential way, one operation at a time. 
It could be a bottleneck for the performable operation of the processor,
so its proper handling  requires special attention when designing
cooperation between the  \gls{PU}s, working in parallel.

 The conclusion that "\emph{the ever-increasing complexity of superscalar processors would have a
	negative impact upon their clock rate, eventually leading to a leveling off of the rate
	of increase in microprocessor performance}" \cite{EPIC:2000} also means, that when introducing 
	a new control layer (\gls{SV}) on top of cores in the processor, its simple combinational logic
    can be operated at a frequency, allowing  high-speed coordination of the cores' work;
    much higher than the clock frequency needed for the cores for
    making sophisticated computations.

\subsubsection{The memory}
The more \gls{PU}s obviously need broader memory access bandwidth,
but the burden of memory bandwidth is not as bad as one might think for the first look.
In the \gls{SPA} systems
one processor is linked trough one bus to one memory decoder.
However, in the hyper-threaded architectures several outstanding memory
access requests can coexist.

One must remember, the producer/consumer model: 
the "memory wall" is still active.
Even with  \gls{EMPA},  the many-core processors
cannot receive more memory contents, than the memory subsystem can produce.
\gls{EMPA} can make, however, good use of multiple memory access devices.
This ability might need to change the memory access architecture:
the many-core processors might need more than one independent memory buses,
the buses can be (time or space) multiplexed, and the memories might need
multiple decoders to the same memory address space.
To broaden the memory-access bandwidth, independent multi-port memories are needed, like \cite{Cypress15}.

In many-processor systems a lot of efforts are needed to provide
coherent operation of the \gls{CPU}s. In \gls{EMPA}, the \gls{PU}s
have \textit{coordinated operation}, so the accidental simultaneous access
can be eliminated by the compiler/\gls{SV}. For examples see 
the sample programs below; the interrupt or \gls{OS} service operation
or direct memory access: the logic of the (cooperative) operation 
excludes the simultanous access, so using the (relatively slow and energy wasting)
shared memories (like in \cite{VishkinHome2007}) are not necessary here.

\subsection{The parent-child relationship}\label{sec:parentchild}
The cores are uniform and independent, but
using the mentioned bitmasks, they can be in parent-child relationship,
to arbitrary depth. A core can have only one parent,
but an arbitrary number of children. This relationship allows for
several generations, unlike the master-slave relationship,
used  in some other architectures like the "Desktop Supercomputer"~\cite{Vishkin:AbstractionCACM}
and allows for dynamic behaviour similar to that of the "Invasive Computing"~\cite{InvasiveComputing:2011}.

Fig.~\ref{fig:EMPAparentchild} attempts to summarize the signal and data
traffic of the cores. At the top of figure the core is in role \textit{parent} and at the bottom in role \textit{child}.
These roles are of course context-dependent, in another context
a child can be parent of another core, or a core cannot have a child at all. The shown storages and signals are typical for that role. 
Akin in \gls{FPGA}s, some well-defined, mostly fixed functionality blocks are
placed in close vicinity to each other, and the end-user has the possibility
to connect them, changing some configuration parameters, which selects
one of the predefined functionalities.

\subsection{The dynamic architecture}
The individual cores take the responsibility for executing a dedicated \gls{QT}.
The \gls{QT}s can be nested, i.e. a core
may face the task to delegate part of its job to another \gls{PU}.
It can be solved (with the active help of \gls{SV})
using the parent-child architecture.
If there is at least one available core, the  \gls{SV} rents it 
from the pool of cores for the requesting core, and administers it as a child of the requesting core, in both cores in the configurable bitmasks.
This means, that the "processing graph" will be mapped to the available cores.
\index{parent-child relationship}
\index{mapping!QTs to cores}

The child core gets enabled and begins its  independent (and parallel!) life. A child core might find a termination metainstruction (in contrast with the conventional processor, where only 'halt' is possible), which leads to notifying the \gls{SV} (see also Figs.~\ref{fig:EMPAparentchild} and~\ref{fig:TwoLevelDiagram}).
The \gls{SV} administers the termination of the parent-child relationship,
and puts back the (former child) core into the pool.
From that moment that core might be rented for another task.

The parent is responsible for performing the complete task of the \gls{QT}
it received, even if it delegated part of the job (in form of child \gls{QT}s)
to its child cores. This means, that a parent \gls{QT} must wait the 
termination of all of its child \gls{QT}s, in order to be sure the work completed. To do so, the \gls{SV} will
block the termination of a parent \gls{QT} until its children mask 
gets cleared.

 \subsection{Data passing} \label{sec:datapassing}
A crucial question is passing data between cooperating cores during processing.
Splitting the code into \gls{QT}s in a reasonable way, allows to
make a bargain between loosing performance because of transferring data
and gaining performance because of using more \gls{PU}s.
Anyhow, \textit{some} data passing in inevitable.

When a piece of the code delegated originally to the parent core
is delegated to the child core, the child core must inherit also the
internal status of the parent, including the contents of the registers.
It needs dedicated wiring between the cores and the \gls{SV},
 and (depending on the
physical location of the cores) can take somewhat longer time than the
other \gls{SV} operations. In this case the synchronization is not a problem: the child core commences its life \textit{after} it received the needed data.

Upon termination of \gls{QT}s, however, synchronization of the eventual returned data might be an issue: the children cannot know
\textit{when} to copy back data into registers of the parent.
To solve this problem, the \gls{SV} latches the data  returned by the child, and when the 
parent is about to terminate or explicitly waits for the 
termination of the child, transfers it to the parent core.
This type of information transfer requires dedicated wiring
between the core and the \gls{SV}, and should be implemented
as a two-stage transfer.
In this case the \gls{SV} acts as a switching center, so making
dedicated wiring between a core and the rest of cores can be avoided.

\newcommand{\SupervisorLevel}{Supervisor level}
\newcommand{\CoreLevel}{Core level}
\newcommand{\Created}{Created}
\newcommand{\Create}{Create}
\newcommand{\Allocated}{Allocated}
\newcommand{\Allocate}{Allocate}
\newcommand{\Deallocate}{Deallocate}
\newcommand{\Enabled}{Enabled}
\newcommand{\Enable}{Enable}
\newcommand{\Disable}{Disable}
\newcommand{\Meta}{Meta}
\newcommand{\Action}{Action}
\newcommand{\Execute}{Execute}
\newcommand{\Executable}{Executable}

\newcommand{\yslant}{0.5}
\newcommand{\xslant}{-0.6}

\begin{figure}
{
\begin{tikzpicture}[scale=1,every node/.style={minimum size=1cm},on grid]

	\begin{scope}[
		yshift=-120,
		every node/.append style={yslant=\yslant,xslant=\xslant},
		yslant=\yslant,xslant=\xslant
	] 
		\draw[black, dashed, thin] (0,0) rectangle (5,6); 
		\draw[fill=blue]  
		    (1,5) circle (.2) 
		    (2,4) circle (.2) 
		    (3,3) circle (.2) 
		    (3,5) circle (.2) 
		    (4,3) circle (.2) 
			; 
		\draw[-latex,thin] 
			(1.2,5) to[out=0,in=90] (2.1,4.2); 
		\draw[-latex,thin] 
			(1.8,4) to[out=180,in=-90] (1.0,4.8); 
		\draw[-latex,thin] 
			(2.2,4) to[out=0,in=90] (3.0,3.2); 
		\draw[-latex,thin] 
			(2.8,3) to[out=180,in=-90] (2.0,3.8); 
		\draw[-latex,thin] 
			(3.2,3.0) to[out=45,in=-45] (3.2,5.0); 
		\draw[-latex,thin] 
			(3.0,4.8) to[out=-90,in=135] (2.8,3.0); 
		\draw[-latex,thin] 
			(3.0,3.2) to[out=90,in=90] (4.0,3.2); 
		\draw[-latex,thin] 
			(4.0,2.8) to[out=-90,in=-90] (3.0,2.8); 
		\fill[black]
		    (1.5,5.4) node[left,scale=.7]{\textbf{\Created}} 
		    (2.5,4.4) node[left,scale=.7]{\textbf{\Allocated}} 
			(1.9,4.9) node [scale=.6, rotate=-40] {\Allocate} 
			(1.1,4.1) node [scale=.6, rotate=-40] {\Deallocate} 
			(2.9,3.9) node [scale=.6, rotate=-40] {\Enable} 
			(2.1,3.1) node [scale=.6, rotate=-40] {\Disable} 
			(3.4,4.6) node [scale=.6, rotate=-40] {\Meta} 
		    (4.6,2.4) node[left,scale=.7]{\textbf{\Execute}} 
			(4.2,3.4) node [scale=.6, rotate=-40] {\Executable} 
		    (3.4,2.3) node[left,scale=.7]{\textbf{\Enabled}} 
			(0.5,6.5) node[right, scale=.7] {\CoreLevel}	
			;	
	\end{scope}	
	\draw[ultra thin](-2,4) to (-2,-0.25); 
	\draw[ultra thin](-0.4,3.6) to (-0.4,-0.45); 
	\draw[ultra thin](1.2,3.5) to (1.2,-0.65); 
	\draw[ultra thin](0.0,5) to (0.0,0.65); 
	\fill[black]
			(-0.75,2.1) node [scale=.6, rotate=90] {cloning, cache sharing} 
			(-0.55,2.1) node [scale=.6, rotate=90] {ForChild,FromParent} 
			(-0.25,2.1) node [scale=.6, rotate=-90] {FromChild,ForParent}
			(-0.05,2.1) node [scale=.6, rotate=-90] {link register}
	;	
	\begin{scope}[
		yshift=0,
		every node/.append style={yslant=\yslant,xslant=\xslant},
		yslant=\yslant,xslant=\xslant
	]
		\fill[white,fill opacity=.75] (0,0) rectangle (5,6); 
		\draw[black, dashed, thin] (0,0) rectangle (5,6); 
		\draw [fill=red]
		    (1,5) circle (.2) 
		    (2,4) circle (.2) 
		    (3,3) circle (.2) 
		    (3,5) circle (.2) 
			;
		\draw[-latex, thin]
			(2.8,5) to[out=180,in=120] (1.8,4); 
		\draw[-latex, thin]
			(2.2,4) to[out=0,in=-60] (3.2,5); 
		\draw[-latex, thin]
			(2.8,5) to[out=-120,in=120] (2.8,3); 
		\draw[-latex, thin]
			(3.2,3) to[out=60,in=-30] (3.2,5); 
		\draw[-latex, thin]
			(3,5.2) to[out=120,in=60] (1,5.2); 
		\draw[-latex, thin]
			(1,4.8) to[out=-60,in=-120] (3,4.8); 
		\fill[black]
			(0.5,6.5) node[right, scale=.7] {\SupervisorLevel}			
			(1.5,5.4) node[left,scale=.7]{\textbf{\Create}} 
			(2.4,3.5) node[left,scale=.7]{\textbf{\Allocate}} 
			(3.5,2.6) node[left,scale=.7]{\textbf{\Enable}} 
			(3.5,5.4) node[left,scale=.7]{\textbf{\Action}} 
			(2.1,5.0) node [scale=.6, rotate=-40] {QxCreate} 
			(2.7,3.4) node [scale=.6, rotate=-40] {QxWait} 
			(2.8,4.0) node [scale=.6, rotate=-40] {QTerm} 	
			; 
	\end{scope} 
\end{tikzpicture}
\caption{How EMPA processor operates at supervisor and core levels}
\label{fig:TwoLevelDiagram}
}
\end{figure}


 \subsection{Two-level operation} \label{sec:twoleveloperation}

Fig. \ref{fig:TwoLevelDiagram}is somewhat similar to a state diagram,
and might help two understand the operation of \gls{EMPA}.
Actually, the actions take place at two different levels.

At the beginning, the \gls{SV} "creates" the cores, i.e. it
initializes its internal data structures and places them 
in a pool of sharable \gls{PU}s. When in the pool, the operation 
of the cores is of course not enabled. One of the cores gets
allocated and will be enabled.

In that state, the core will work as a traditional processor,
with the exception that the \gls{SV} can disable its operation,
and during the pre-fetch stage it decides whether the next instruction
is a normal executable instruction or a meta-instruction.
In the former case, the core executes the executable instruction
as the conventional processors do.
In the latter case, using its 'Meta' signal, the core notifies 
\gls{SV}. In response, the \gls{SV} takes over the execution
of the metainstruction: advances the \gls{PC} of the core
to the next instruction at the core level, and 'executes'
the meta-instruction at the supervisor level.

If the metainstruction is to create a \gls{QT}, it means that
new core(s) must be rented from the pool (the \gls{HW} is provided),
 and equipped with proper internals (the \gls{SW} is provided). 
Then the new core begins its independent life. When it reaches 
a metainstruction, it notifies the \gls{SV}.

Waiting can occur only as an effect of a metainstruction.
When there are no more cores in the pool at the moment, or a parent is about
to terminate without its children being terminated or an explicit
waiting requested, the \gls{SV} simply disables the core,
until the condition fulfilled.

Parameter passing also happens under \gls{SV} control.
The latched registers (to/from child/parent) are filled in the partners correspondingly, when creating or terminating a \gls{QT}
(executing \lstinline|QxCreate|
or \lstinline|QTerm|) triggers the action. As shown,
\gls{SV} is fully responsible for syncronized data transfer between latched registers,
but its operation is based on how the programmer configures 
\gls{SV} through issuing the corresponding metainstructions.

 \subsection{Pseudo registers} \label{sec:pseudoregisters}
 Another crucial question is how such unusual data passing can be
 used ("programmed") in a way, close to the conventional one.
 The conventional cores use registers for quick manipulation of data,
 so a useful idea is  to use pseudo-registers for transferring
data between a child and its parent.
In this way both the parent and the child see a "register", which is
a well-known term for the cores. However, it is not an item in the
register file: it has a register address, but it also has a
context-dependent functionality.
In this way the parent and the child can share some data
in a way, which appears as handling registers.

This pseudo-register migh have a bit longer access time
(depending on the physical position of the cores and the additional electronic
functionality, hidden under the facades of the register and depending on the operating mode),
but surely shorter than reaching any memory or using any kind of internal network.
Different operating modes for collective work can be defined,
and through those (pseudo)registers  the quickest possible
data transfer can be reached.

In normal mode a pseudoregister behaves (nearly) as a traditional register,
except that it is mapped to a latch register of the core.
In mass processing mode  a pseudo register behaves in a quite special way. For example, what the parent writes in its own register,
children can read from their own register, and what children write in their own register,
the parent can read from their own register. In order to avoid syncronization issues,
the \gls{SV} latches the sent data when the sender is ready
to send it, and allows the receiver to read the data from the 
latch when the receiver is ready to accept it.

As shown in Figs.~\ref{fig:EMPAparentchild} and \ref{fig:TwoLevelDiagram}, (pseudo)register transfer
might happen in several situations, and in all cases under the control
of the \gls{SV}.
The first data transfer occurs when renting a new core. 
First of all, the \gls{SV} handles the corresponding bitmasks 
'Parent' and 'Children', and clones the complete internal state
(including the register file and the \gls{PC})
of the parent to the new child. When ready, the \gls{SV} 
enables the child, which begins its independent (and parallel!)
life with executing the delegated code chunk.
After this, the parent skips that (logically already executed) code fragment
and continues execution at that new address.
When the child is about to terminate, it notifies its \gls{SV}
with the 'Meta' signal. In response, the \gls{SV} latches
the content of the link register of the child core for the parent,
changes the contents of register 'Children' and 'Parent' correspondingly. The latched register content is available
for the destination parent core through issuing an explicit or implicit 'Wait' (recall that  termination implies a 'Wait'),
and will be written from the latch into the corresponding register 
only when the parent requests so. In this way no syncronization
issues might happen.

When cores are allocated for mass processing, registers 'ForChild' and 'FromChild' 
will be initialized in the parent core. The parent can write 'ForChild'
through writing its own pseudoregister and read 'FromChild' through
reading its own pseudoregister. (Remember, the \gls{PC} of the parent
might stall at the address whare mass processing begins; and also
that \gls{SV} can read/write any of the registers, independently
of the operation of the  cores.)

In mass processing mode, the 'Mode' code and 'ForChild' are latched by the child
when preallocatin cores. 
If a child writes to its own pseudoregister,
the value will be latched in 'FromChild' in the parent, and the next (repeated)
\gls{QT} creation might consider that value.
Depending on the 'Mode', the parent might watch and read 
through reading its own pseudoregister
\index{register!latched for another core}
the latched value written by the child to its own pseudoregister.
Since for reading and writing pseudoregister as 'parent' or as 'child' there are two different directions
(two different latched registers), some special rules determine its context-dependent utilization, see \cite{vegh2}.
To forward data (i.e. to transmit, data received on its input,
to its output) the core needs to use an explicit copying from the
input pseudoregister to the output pseudoregister instruction.
For the programming implementation see \cite{vegh2}.

\MESourceFile[language={[y86]Assembler}, wide, options={numbers=left},number=-1,
] {Qasum0_4.Eyo}
{The sum-up routine in traditional coding}{lst:Qasum04}

\section{Utilizing the architecture} \label{sec:utilizingempa} 
After having an architecture presented above, one shall find utilization
possibilities, and implement them. As an example of the possibilities,
a program summing up elements of a vector is presented.
The example in Listing~\ref{lst:Qasum04} has been adapted from the 'asumup' program~\cite{hallaron}.
The program is written in Y86 assembly language, extended with \gls{EMPA} metainstructions.
The toolchain for  \gls{EMPA}
(including the assembler and simulator, as well as some programming methodologies)
is described in a separate paper~\cite{vegh2}.
Since Y86 is a simplified for education version of the widely known Intel x86 processor,
the coding can be followed easily. Here only a qualitative description
can be given. More programming details are described in \cite{vegh2},
the mentioned sample programs, toolchain, their user guides 
are available from~\cite{veghEMPAthY86:2016}.

First the number of arguments and vector address (lines 3-4)
are loaded into registers.
Then the program clears the sum (line 6),
and  verifies the number of items (line 8).
The kernel of the calculation is in lines 9-15. 
First the actual vector element is loaded 
into \lstinline|%esi|,
then  it is added 
(lines 9-10)
to the partial sum stored in \lstinline|%eax|.
After this, the address of the actual element is advanced properly (lines 11-12), 
then the loop counter advanced (lines 13-14).
.
Finally, depending on the actual conditions (line 15)
 the loop
returns back to repeat the procedure for the next vector element.

\subsection{Eliminating obsolete instructions'(FOR)}
As it can be seen from Listing~\ref{lst:Qasum04}, this implementation is 
rather ineffective:  the payload work is done in lines 9-10,
while lines 11-15 only serve loop organization. If the \gls{SV}
could take over loop organization, we could reach a performance 
gain about 3. This needs thinking in terms of \gls{EMPA}.

Actually, the program fragment needs the address of the current summand
and a register to store the partial sum. I.e. we can separate lines 9-10
in a \gls{QT} and organize the job around this approach.
So, lines 9-10 will be executed by the child, on the request from the 
parent. As mentioned, the parent clones its "glue" to the child,
so the address of the array is accessible for the child in \lstinline|%ecx|,
the old partial sum is delivered in \lstinline|%eax|, and the new
partial sum is cloned back to the parent also in \lstinline|%eax|.
The child terminates after executing lines 9-10, allowing the parent
to know that one element has been added to the partial sum.
So, the job for providing the right address and counting the iterations
remains for the parent. Since the parent is only waiting while the 
child terminates, its arithmetic facilities can be used for this task.

The parent pre-allocates a child for the work. The pre-allocation
is needed because the other cores might allocate cores in parallel,
so through the preallocation is guaranteed that always will be available
core for the iterations. That child core will be allocated by the parent as many times as needed,
and so the loop will be jointly executed by the child and the parent.
The parent starts executing an iteration (as many times as needed),
and waits until the child terminates. The \gls{SV} also participates 
in the game: calculates the address of the vector element for the next iteration
and delivers data and signals between parent and child.

The parent writes the current address to its latched 'ForChild' register
(see Fig.~\ref{fig:EMPAparentchild})
and it will be latched by the child upon creation into its own the latched 'FromParent'
register. The number of the remaining iterations are stored in the latched
'FromChild' in the parent. Before instructing the parent to start a new iteration, 
\gls{SV} checks if the content of 'FromChild' is cleared. If not, 
it instructs the parent to start the next iteration and decrements 
the latched 'FromChild' in the parent. Since the child can write its latched
'ForParent' register, the content of which is transferred upon termination
to the latched 'FromChild' in the parent, the child can \lstinline|break|
the loop.

\subsection{Eliminating obsolete stages (SUMUP)}
Summing up elements in a vector is a simple and ideal example of processes
which cannot be parallelized: in the addition one of the summands is the
previous partial sum, so the next summing cannot be started until the
previous one is terminated.

The closer look shows that it is because in the frame of executing
a single instruction the processors must read the content of a register and write it back,
updating it with the new partial sum. We can also see that the partial
sum is never used, we are only interested in the final sum.
This means, that \textit{if we can find a way in the architecture, which allows a
cooperative execution of adding, furthermore a separated readout or
the final sum, we can parallelize this strictly sequential process}.

To implement this, needs a bit more functionality from the \gls{SV}.
Suppose we have enough cores which can be allocated, and preallocate them.
In this special operating mode which allows opening execution stages 
for a child, an adder is prepared in the parent, which on one input
receives the latched 'FromChild', and on the other input receives
its own output (the previous partial sum). 
The child receives the current address as described in the previous section,
and in this special mode executing \lstinline|addl| to a special 
pseudo register means writing to 'ForParent' in the child, which also
triggers transferring to 'FromChild' in the parent, i.e. the content
read out by the child will be added to the partial sum stored in the parent.
%

\subsection{System services and other uses}

Some system services, for example semaphore handling, do not really need
all the facilities of the \gls{OS}, so they can be implemented in alternative way.
As our former measurements on soft system~\cite{Vegh:2014:ICSOFTsemaphore} proved,
such alternative implementation resulted in performance gain about 30,
although in that case no context changing was needed.
Similar gain can be expected when implementing \gls{OS} services with \gls{EMPA}.
The gain factor will surely be increased because of the eliminated 
context change, but the concrete gain will depend on the functionality of the service.

\section{Performance of the accelerator}\label{sec:performanceempa}
As discussed above, using \gls{EMPA} architecture can distribute
the code between \gls{PU}s,
can eliminate
obsolete instructions, obsolete execution stages, obsolete context changes, etc.
The machine instruction execution remains the same, so to measure performance
is not simple at all. Practically, the speedup (the ratio of the execution times)
remains the only measurable quantitity. Recently, a figure of merit characterizing
the \textit{effective parallelization}~\cite{Vegh:AmdahlMerit:2016} has been developed,
so below that merit will be used to describe the performance gain due to utilizing \gls{EMPA},
and it will be compared to the traditionally used merit: the speedup 
divided by the number of cores ($\frac{S}{k}$).
The effective parallelization can be derived from the number
of \gls{PU}s $k$ and the measured speedup $S$ as

\begin{equation}
\alpha_{eff} = \frac{k}{k-1}\frac{S-1}{S} \label{equ:alphaeff}
\end{equation}

The conventional methods of parallelization
suffer from inefficiency in using computing power of multiple \gls{PU}s: 
because of the presence of the sequential-only part,
the more cores are used, the lower is the value of $\frac{S}{k}$,
while $\alpha_{eff}$ really describes correctly how effectively
the cores are utilized.  The 'sumup' program has been tested in three versions: 
using the conventional programming (i.e. NO \gls{EMPA} acceleration),
replacing "control" instructions with \gls{SV} activity in the FOR mode,
and using SUMUP mode, where (in addition to eliminating control instructions) the cooperation eliminates
the unneeded  read/write back cycles \textit{within} a machine instruction.

The results (measured using the simulator~\cite{veghEMPAthY86:2016}) are compared in Table~\ref{tab:effectivealphaEMPA}.
for different vector length and different number of cores.
The simulator uses arbitrary, but reasonable
execution times, expressed in units of the control clock driving 
the \gls{SV}. The actual values might change when an electronic
version (\gls{RC} implementation) allows to provide more accurate data.

\MEtable{
\begin{tabular}[resize]{rrrclll} %
 \rowcolor{tableheadcolor}Vector &Mode of & Time	    & No of      & Speedup &  $\frac{S}{k}$& $\alpha_{eff}$\\
 \rowcolor{tableheadcolor} length & mass proc & (clocks)	& cores (k)  &   (s)   &               &\\
 \midrule
1 & NO    & 52  &   1  &  1    &   1   &  1 \\
1 & FOR   &31   &   2  & 1.68  &  0.84 &  0.81 \\
1 & SUMUP &33   &   2  & 1.58  &  0.79 &  0.73 \\
 \midrule
2 & NO    & 82  &   1  &  1    &   1   &  1 \\
2 & FOR   & 42   &   2  & 1.95  &  0.98 &  0.97 \\
2 & SUMUP & 34   &   3  & 2.41  &  0.80 &  0.87 \\
 \midrule
4 & NO    & 142  &   1  &  1    &   1   &  1 \\
4 & FOR   &64   &   2  & 2.22  &  1.11 &  1.10 \\
4 & SUMUP &36   &   5  & 3.94  &  0.79 &  0.93 \\
 \midrule
6 & NO    & 202  &   1  &  1    &   1   &  1 \\
6 & FOR   & 86   &   2  & 2.34  &  1.17 &  1.15 \\
6 & SUMUP & 38   &   7  & 5.31  &  0.76 &  0.95 \\
\end{tabular}
}{Effective parallelization in EMPA processor, in different operating modes}{tab:effectivealphaEMPA}{}{}

\MEtikzfigure{
		\begin{tikzpicture}
		\begin{axis}[
		/pgf/number format/.cd,
		use comma,
		1000 sep={},
		legend style={
			cells={anchor=east},
			legend pos=north west,
		},
		xmin=0, xmax=7,
		ymin=1, ymax=5.5, 
		xlabel=Vector length,
		ylabel=Measurable speedup,
		]
		\addplot[ very thick, color=red,mark=*] plot coordinates {
		 (1,1.68)
		 (2,1.95)
		 (4,2.22)
		 (6,2.34)
		};
		\addlegendentry{FOR performance gain}
		\addplot[ very thick, color=blue,mark=*] plot coordinates {
		 (1,1.58)
		 (2,2.41)
		 (4,3.94)
		 (6,5.31)
		};
		\addlegendentry{SUMUP performance gain}
		
		\end{axis}
		\end{tikzpicture}
}
{Diagram showing data from Table 
 \protect{\ref{tab:effectivealphaEMPA}}: 
 the measurable speedup for two different mass processing methods, in function of the vector length.}
{fig:Speedup}{}{}

\subsection{Analyzing speedup results}
As Table \ref{tab:effectivealphaEMPA} shows, both the conventional
and \gls{EMPA} execution times increase linearly with the length
of the vector. Their intersect and slope values, however,
are very different.

The FOR mode of \gls{EMPA} is nearly
3 times quicker than the conventional method,
as some computed control statements are replaced by the
much more effective \gls{SV} functionality.
This requires only 2 \gls{PU}s.

In the SUMUP method, in addition to omitting the
computed control machine instructions, even the obsolete
fetch, decode, writeback, etc. stages of one instruction execution
in the loop kernel
are replaced by  \gls{SV} functionality.
For the SUMUP
mode an extra element increases the execution time by one clock cycle,
at the price of utilizing one more \gls{PU}.
This behavior is especially valuable, because using conventional
methods of parallelization the algorithm cannot be parallelized at all.

The measured speedup values are derived from a mixture of different types of instructions: both conventional and \gls{EMPA} code contains both sequential and parallel parts, so despite of the linear 
increase, the measured speedup will not linearly depend on the 
vector length, see Fig. \ref{fig:Speedup}. The two speedup values
will saturate for high vector lengths at values $\frac{30}{11}$
and $30$, respectively.

\subsection{Analyzing parallelization efficiency}

When eliminating
in this very simple loop the control instructions in FOR mode,
 the $\frac{S}{k}$ values can even be \textit{above} unity. 
This means \textit{not} a higher \gls{PU} performance,
it is due to the more clever organization of cycles.

\MEtikzfigure{
		\begin{tikzpicture}
		\begin{axis}[
		/pgf/number format/.cd,
		use comma,
		1000 sep={},
		legend style={
			cells={anchor=east},
			legend pos=north west,
		},
		xmin=0, xmax=7,
		ymin=.7, ymax=1.25, 
		xlabel=Vector length,
		ylabel=Core utilization efficiency,
		]
		\addplot[ very thick, color=red,mark=*] plot coordinates {
		 (1,0.84)
		 (2,0.98)
		 (4,1.11)
		 (6,1.17)
		};
		\addlegendentry{FOR $\frac{s}{k}$}
		\addplot[ very thick, color=red,mark=x] plot coordinates {
		 (1,0.81)
		 (2,0.97)
		 (4,1.10)
		 (6,1.15)
		};
		\addlegendentry{FOR $\alpha*{eff}$}
		\addplot[ very thick, color=blue,mark=*] plot coordinates {
		 (1,0.79)
		 (2,0.80)
		 (4,0.79)
		 (6,0.76)
		};
		\addlegendentry{SUMUP $\frac{s}{k}$}
		\addplot[ very thick, color=blue,mark=x] plot coordinates {
		 (1,0.73)
		 (2,0.87)
		 (4,0.93)
		 (6,0.95)
		};
		\addlegendentry{SUMUP $\alpha*{eff}$}
		
		\end{axis}
		\end{tikzpicture}
}
{Diagram showing data from Table 
 \protect{\ref{tab:effectivealphaEMPA}}: 
 the core utilization efficiency for two different mass processing methods, in function of the vector length.}
{fig:UtilizationEfficiency}{}{}

In the SUMUP mode, the helper cores are utilized only for a short
period of time, so the utilization efficiency is low for short vectors.

Note that since the \gls{PU}s
are put back in the pool, for very long vectors much lower number of cores might be needed.
If the compiler can find out the length of processing in that mode
(in our example it is 30 clock cycles), it should not allocate 
more that that number of cores: when the parent needs the 31st core,
the 1st core is available again, so the summing can be continued
for arbitrary vector length.
In calculating the effective utilization of cores using Equ. \ref{equ:alphaeff},
$k$ should be replaced with 
$\max\limits_{k\leq  k_{eff} \leq 30}$.

\MEtikzfigure{
		\begin{tikzpicture}
		\begin{axis}[
		/pgf/number format/.cd,
		use comma,
		1000 sep={},
		legend style={
			cells={anchor=east},
			legend pos=north west,
		},
		xmin=0, xmax=600,
		xmode=log,
		ymin=.45, ymax=1.15, 
		xlabel=Vector length,
		ylabel=Core utilization efficiency,
		]
		\addplot[ very thick, color=blue,mark=*] plot coordinates {
		 (2,0.79)
		 (3,0.80)
		 (4,0.80)
		 (4,0.79)
		 (5,0.79)
		 (6,0.77)
		 (11,0.70)
		 (16,0.63)
		 (21,0.57)
		 (26,0.52)
		 (31,0.48)
		 (51,0.60)
		 (101,0.74)
		 (201,0.84)
		 (501,0.91)
		};
		\addlegendentry{SUMUP $\frac{s}{k}$}
		\addplot[ very thick, color=blue,mark=x] plot coordinates {
		 (2,0.73)
		 (3,0.88)
		 (4,0.92)
		 (5,0.93)
		 (6,0.94)
		 (11,0.96)
		 (16,0.96)
		 (21,0.96)
		 (26,0.96)
		 (31,0.96)
		 (51,0.98)
		 (101,0.99)
		 (201,0.99)
		 (501,1.00)
		};
		\addlegendentry{SUMUP $\alpha*{eff}$}
		
		\end{axis}
		\end{tikzpicture}
}
{Efficiency $\frac{S}{k}$ and $\alpha_{eff}$ for EMPA processor in the SUMUP mode, in function of the vector length.}
{fig:SaturationionEfficiency}{}{}

The two different points of view of the two merits is best reflected in Fig. \ref{fig:SaturationionEfficiency},
where $\frac{S}{k}$ and $\alpha_{eff}$ are shown in function of the vector length,
using an \gls{EMPA} processor in SUMUP mode.
Because of the effect of sequential code fragments, both curves
start increasing with increasing the number of the cores.
As mentioned, 
in this mass processing mode max. 31 cores (1 parent plus 30 child cores) can be used.
There is no sense to use more then 30 child cores: they would need to wait for sending 
their summand for the parent.
Because of this, both the 
number of threads and the speedup keep raising, while the number of the cores
saturates at 31. For short vectors, $\alpha_{eff}$  is relatively low,
because the helper cores are used only in a fragment of time.
As all the 30 helper cores have "full time job", the $\alpha_{eff}$
dependence saturates at value $1$. In contrast, $\frac{S}{k}$ starts
to decrease with increasing the number of the cores,
and after reaching 30 cores, the speedup continues, but $k$ remains constant, so the dependence turns back and saturates also at value $1$, but approaches it much more slowly.

\section{Conclusions} \label{sec:conclusions} 

The idea of introducing \gls{EMPA} in processor technology opens
a series of new possibilities. As main accelerator, it allows to turn 
a many-core processor to an extremely high-power single-core processor.
To make an old single-thread program many-core aware, it is enough to
recompile the program using an \gls{EMPA}-aware compiler and run it 
on an \gls{EMPA} architecture processor.
\gls{EMPA} uses no hidden \gls{PU}s: the same cores can be "rented" for normal
code execution and out-of-order or speculative evaluation.
This means that the superfluous logic~\cite{EPIC:2000} concerting and hiding the extra \gls{PU}s
can be omitted, simplifying the internal architecture, reducing the number of
transistors and reducing also the power consumption.

For calculational applications, several hundred times higher
parallelization can be reached: the compile-time discovery of parallelization 
possibilities and mixing thread and instruction level parallelism 
tends to reach the ideal case of "infinite resources"~\cite{NicolauFischer1984}.
In addition to that theoretically checked possibility, the \gls{SV} can 
more efficiently perform some control functions from loop organization
to opening the closed von Neumann execution frames for the helper cores,
raising at least one more order of magnitude in the performance.

Using \gls{OS}s are getting more simple and more effective with
using \gls{EMPA}: no context changing is needed, and the user mode and 
kernel mode programs can run in (at least partly) parallel.
Since \gls{QT}s are by their nature atomic processing units, a big part of  
operating systems dealing with semaphors for shared resource usage,
scheduling, etc. becomes obsolete (greatly reducing the amount of codes,
both writing and testing),
and also the built-in synchronization of \gls{EMPA} can replace those
services offered by the \gls{OS}s.

The real-time characteristics of processors also benefit from \gls{EMPA}.
To service an interrupt, no state saving and restoring is needed,
saving memory cycles and code. The program execution will be predictable: 
the processor need not be stolen from the running main process.
The atomic nature of executing \gls{QT}s will prevent issues like
priority inversion, eliminating the need for special protection protocols.

From the point of view of accelerators, an \gls{EMPA} processor 
provides a natural interface for linking special accelerators to the
processor. Any circuit, being able to handle data and signals shown in
Fig.~\ref{fig:EMPAparentchild} can be linked to an \gls{EMPA} processor with easy.

\bibliographystyle{IEEEtran}
\bibliography{IEEEabrv,Bibliography}
%
%
%

%

\begin{IEEEbiography}[{\includegraphics[width=1in,height=1.25in,clip,keepaspectratio]{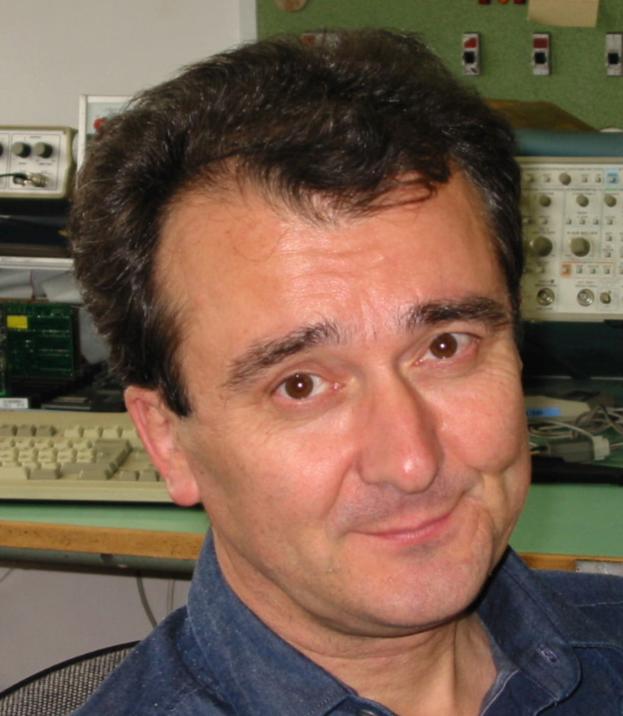}}]{V\'EGH, J\'anos}
received his PhD in Physics in 1991, ScD in 2006.
He is working in computing since the early 80's, and since 2008
is a full professor in Informatics, currently with University of Miskolc, Hungary.
In research, he is looking for ways to prepare more performant and 
more predictable computing, especially using mainly multi-core processors.
He is also dealing with soft processors and hardware-assisted operating 
systems, reconfigurable and many-processor computing. 
\end{IEEEbiography}




\end{document}